\documentclass[preprint]{aastex7}

\linenumbers

\newcommand{\Msun}{\ifmmode {M_{\odot}}\else${M_{\odot}}$\fi}
\newcommand{\lessim }{{\lower0.8ex\hbox{$\buildrel <\over\sim$}}}
\def\amin{\ifmmode^{\prime}\else$^{\prime}$\fi}
\def\asec{\ifmmode^{\prime\prime}\else$^{\prime\prime}$\fi}

\newcommand{\be}{\begin{equation}}
\newcommand{\ee}{\end{equation}}
\newcommand{\nn}{\mbox{} \nonumber \\ \mbox{} }
\newcommand{\ba}{\begin{eqnarray}}
\newcommand{\ea}{\end{eqnarray}}
\newcommand{\om}{\omega}

\newcommand{\E}{{\bf E}}
\newcommand{\B}{{\bf B}}
\newcommand{\J}{{\bf J}}
\newcommand{\A}{{\bf A}}
\renewcommand{\v}{{\bf v}}
\renewcommand{\k}{{\bf k}}
\newcommand{\x}{{\bf x}}
\newcommand{\p}{{\bf p}}
\renewcommand{\H}{{\cal H}}

\newcommand{\Bf}{{magnetic field}}
\newcommand{\Bfs}{{magnetic fields}}
\newcommand{\NS}{neutron star}
\newcommand{\WD}{white dwarf}

\newcommand{\ms}{magnetosphere}
\newcommand{\NSs}{{neutron stars}}

\newcommand{\Ef}{{electric  field}}

\newcommand{\EM}{electromagnetic}

\newcommand{\mss}{magnetospheres}

\newcommand\etal{\it{et al.\ }}
\newcommand\eg{{\it{e.g.\ }}}

\begin{document}

\title{Long-period radio transient PSR J0901$-$4046 is not an Isolated White Dwarf Pulsar}

\author[orcid=0000-0001-6436-8304,sname='Lyutikov']{Maxim Lyutikov}
\affiliation{Department of Physics and Astronomy, Purdue University, 525 Northwestern Avenue, West Lafayette, IN 47907-2036, USA}
\email[show]{lyutikov@purdue.edu}  

\author[orcid=0000-0001-6098-2235,sname='Kilic']{Mukremin Kilic} 
\affiliation{Homer L. Dodge Department of Physics and Astronomy, University of Oklahoma, 440 W. Brooks St., Norman, OK, 73019 USA}
\email{kilic@ou.edu}

\author[orcid=0000-0003-1915-5670, sname='Wolszczan']{Alexander Wolszczan}
\affiliation{Department of Astronomy and Astrophysics, Pennsylvania State University, 525 Davey Laboratory, University Park, PA 16802,
USA}
\affiliation{Center for Exoplanets and Habitable Worlds, Pennsylvania State University, 525 Davey Laboratory, University Park, PA 16802,USA}
\email{aleks1853@gmail.com}

\author[orcid=0000-0003-3944-6109,sname='Heinke']{Craig O. Heinke}
\affiliation{Physics Department, CCIS 4-181, University of Alberta, Edmonton, AB T6G 2E1, Canada}
\email{heinke@ualberta.ca}

\begin{abstract}
We report the {\it Chandra} non-detection of PSR J0901$-$4046, a $P=75.89 $ seconds long-period radio transient (LPT). For a distance of 467 pc, the upper limit on X-ray luminosity is $L_X \leq$ few $\times 10^{28}$  erg s$^{-1}$. For the  measured $P$ and $\dot{P}$,  this upper limit,  approximately 50 times lower than the previous {\it Swift} observations, is comparable to the spin-down luminosity of a neutron star, but would be approximately four orders of magnitude smaller than the spindown power of a white dwarf.  Our results disfavor isolated WDs as the central star in  PSR J0901$-$4046. We suggest  that the isolated LPTs are powered by magnetic dissipation (not rotation), in a way similar to magnetars' radio emission.

\end{abstract}


\section{Introduction}

Long-period radio transients  (LPTs) are sources of coherent radio pulses with periods ranging from tens of seconds to tens of minutes \citep{Hurley-Walker2022Natur.601..526H,Hurley-Walker2023Natur.619..487H,2022NatAs...6..828C,Caleb2024NatAs...8.1159C,2025ApJ...982L..53W,2025NatAs...9..393L,2026JHEAp..5200566R}. They consist of two distinct  classes: (i)  accretion-powered binary systems with a white dwarf companion, (ii) more mysterious isolated sources. Both types are often called pulsars, mostly by analogy with periodic emission  from rotationally-powered pulsars, but the physical origins of emission in the two cases are likely dramatically different. 

In case (i), 
a class of {\it accretion-powered} ``White Dwarf Pulsars'' has been established over the last decade, AR Sco 
being  the first and most famous system \citep{marsh16,buckley17}. AR Sco
is a 3.56 hour period white dwarf + M dwarf binary at a distance of 116 pc, where the white dwarf has a spin period of 117 s, $\dot{\nu}=-5\times10^{-17}$ Hz s$^{-1}$, inferred spin down luminosity of $2\times10^{33}$ erg s$^{-1}$, and x-ray luminosity of $L_X\sim4\times10^{30}$ erg s$^{-1}$ \citep{gaibor20,takata21,pelisoli22}. 
The spin-down power is an order of magnitude larger than the electromagnetic radiation from the system.
Part of the x-ray luminosity is produced at the surface of the donor star, which is not present for an isolated white dwarf pulsar. However,
part of the emission is modulated at the spin period, and therefore comes from the white dwarf's magnetosphere. 

Other  systems that can be classified as accretion-powered white dwarf pulsars include J191213.72$-$441045.1 
\citep{pelisoli23}, dubbed an AR Sco twin. In this  white dwarf + M dwarf binary with a 4.03 hour period, 
a 5.3 min pulsed emission is detected from radio to x-rays. In a follow-up study, \citet{pelisoli24} did not find any radio emitting white dwarfs ($\gtrsim$1-3 mJy) outside of interacting binaries in the Very Large Array Sky Survey \citep{gordon21}. 
In addition, \citet{zhang25} searched for coherent radio signals from five isolated, rapidly rotating, and magnetized white dwarfs, but
did not detect any emission down to $\mu$Jy levels.
Hence, we now know two binary white dwarf pulsars, but no rotation-powered (isolated) ``White Dwarf Pulsar''.

The origin of the second type of long-period radio transients, likely isolated, is more questionable.  The key point is that such slowly-rotating regular \NSs\ do not  have enough electric potential to break vacuum and generate a flow of relativistic pairs, thought to be needed for the generation of pulsar-like radio emission \citep{GJ,RS75,AronsScharlemann}, see \S \ref{isolatedWD}.
To compensate for slow rotation, {\it super-magnetar} \Bfs\ would be required \citep{2023MNRAS.520.1872B,2023PhRvD.107h1301S,2023ApJ...943....3T}.   But the 
longest period case,  the 6.45-hour  ASKAPJ1839-0756 \citep{2025NatAs...9..393L}, would likely make it unrealistic even for magnetar-scale fields to achieve vacuum breakdown (spindown has not been measured here).

An alternative possibility is that the second type of long-period radio transients are isolated white dwarfs. Magnetospheric emission is possible for isolated WDs,  as initially argued by    \citet{usov88}; see \S \ref{isolatedWD}.

Detection of non-thermal emission from white dwarfs was long expected, by analogy from other compact objects -  neutron stars. 
Neutron stars produce  non-thermal emission via two very different channels: (i)  rotation-powered, mostly isolated neutron stars - pulsars; (ii) accretion-powered (mostly X-ray pulsars). These  are  two distinct  astrophysical phenomena, though there is also a cross-over class of ``transitional pulsars'', which show  changes from rotation-powered to accretion powered states \citep{2022ASSL..465..157P}. An exciting possibility is a similar dichotomy of rotation-powered/accretion-powered systems translated to white dwarfs.

\section{The case for isolated White Dwarf Pulsars}
\label{isolatedWD}

The existence of rotation-powered  white dwarf pulsars has been predicted more than 30 years ago \citep{usov88}. At the core is a possibility of particle acceleration and pair production by the rotation induced \Ef, similar to the case of \NSs\ \citep{GJ}. Early models of pair production in the \NSs' \mss\ were based on the  curvature emission processes.  It is recognized that two-photon pair production can play an important role \citep[e.g.,][]{philippov18}. In the case of white dwarfs with large radii and small \Bfs, it is the  two-photon pair production that may be dominant over curvature emisison.

White dwarf pulsars have been hypothesized to provide explanation for some exotic $\gamma$-ray and radio sources \citep{usov93,zhang05}, and to contribute to the $e^{\pm}$ cosmic ray background \citep{kashiyama11}. However, no pulsed nonthermal emission from any isolated white dwarf has been discovered so far \citep{pelisoli24}, neither in radio nor in X-rays.

A key prediction of rotation-powered white dwarf emission is the production of high energy emission. In case of pulsars, the radio, high energy and very high energy detections are highly overlapping. The basic reason is that vacuum break-down is always accompanied by  high energy emission. 

An important astrophysical question is whether an isolated white dwarf can produce nonthermal emission - a true, isolated  {\it White Dwarf  Pulsar}. The known population of white dwarfs includes a significant fraction of high-mass ($M>1~M_\odot$) products of binary mergers  \citep{cheng20,temmink20,jewett24}, which, because of their evolutionary history, tend to be rapid rotators \citep{schwab21}, and to possess high magnetic fields \citep{garciaberro12}. These stars can, in principle, emit coherent radiation in a way 
analogous to the (still unknown) pulsar emission mechanism \citep{usov88}. 
\citet{zhang25} performed a sensitive radio search for periodic pulses from five massive, magnetic, and rapidly rotating isolated
white dwarfs, and ruled out emission down to $\mu$Jy levels. However, given the relatively small sample size and the 
typical beaming fractions of $\sim 5-20$\% in radio pulsars \citep{2007MNRAS.380.1678K}, geometric effects alone can account for the non-detection of radio pulses
if the emission is narrowly beamed. Based on their null result, \citet{zhang25} conclude that for the isolated white dwarf pulsar hypothesis, either the radio emission is narrowly beamed or intrinsically weak in these systems.   

The recent discovery of an extremely slow pulsar, PSR J0901$-$4046, with a 75.9 s period and surface field of $1.4 \times 10^{14}$ G \citep{2022NatAs...6..828C}, renews interest in the possibility of isolated white dwarfs producing coherent radio emission. The timing analysis rules out binarity in PSR J0901$-$4046. The discovery of such a slow pulsar is a truly exceptional, but puzzling result; the requirement of a potential drop, $\Phi > 10^{12}$ V, needed for the $e^{\pm}$-pair
production and generation of the coherent radio emission \citep{ruderman75} is clearly violated for this pulsar, which has $\Phi \sim 2 \times 10^{11}$ V.  PSR J0901$-$4046 is clearly unusual among the radio emitting neutron star population: this indicates a possible existence of a new class of previously unobserved, very slowly rotating pulsars.

Theoretically, there is a   concept of a death line in pulsars, see Fig. 1 in \citet{2022NatAs...6..828C}.
In order to break vacuum (and generate coherent emission) the potential drop of a relativistic \ms,
\be
\Phi _{\ast} \sim \frac{\mu}{2}  \left( \frac{  \Omega }{c}  \right) ^2, \, 
\mu \sim B_\ast R_\ast^3
\label{Phimu}
\ee
($\mu$ is magnetic moment of the star)
should be sufficiently high, $\Phi \geq 10^{12}$ V.

When expressed in terms of the observed period $P$ and period derivative $\dot{P}$, the potential is
\be
\Phi _{\ast} \sim 2 \pi \frac{e}{\sqrt{c}} I_\ast^{1/2} P^{-3/2} \dot{P}^{1/2}
\ee
where $I_\ast$ is the moment of inertia. 
If J0901-4046 is a \NS,  the resulting potential is
$
\Phi _{NS} \approx  3 \times 10^{11} \, {\rm V}
$, well below the pulsar death line. 
On the other hand, since moments of inertia of {\WD}s  are much larger than those of \NSs, $I_{WD} \sim 10^5 I_{NS}$, for the same measured  $P$ and $\dot{P}$ a \WD\ has $\sim$ a few hundred times larger potential, $ \Phi _{WD} \approx 10^{14} \, {\rm V} $, well above the death line. Thus, given the observed $P$ and $\dot{P}$, a \WD\ will 
much more easily 
produce pairs, and coherent radio  emission.

Coherent emission in pulsars is associated with  pair production and  is usually accompanied by X-ray emission.
It is expected that the X-ray luminosity is a fraction of spin-down power $L_{sd}$
\be
L_{sd} \sim \frac{\mu ^2 \Omega^4}{c^3} \approx \frac{(2\pi)^2 \dot{P} } {P^3} I_\ast
\label{Lsd}
\ee
Equation (\ref{Lsd}) clearly indicates that for a given observed period $P$ and period derivative $ \dot{P} $, a \WD, with a much larger moment of inertia, would have a much larger spin-down luminosity than a \NS, by a factor $\sim 10^5$. In the case of J0901$-$4046, the corresponding estimates are $L_{sd, NS} \sim 10^{28} $ erg  s$^{-1}$ and $L_{sd, WD} \sim 10^{33} $ erg  s$^{-1}$. The accelerating potential is correspodingly larger as well, $\Phi \propto \sqrt{L_{sd}}.$

Typically, a pulsar's X-ray emission is a small fraction of its spin-down power \citep{possenti02}. Though the spread in efficiencies is large,
values of $L_x \sim 10^{-3} L_{sd}$ are typical among millisecond pulsars \citep{bogdanov06}. For comparison, AR Sco has $L_{sd} = 2\times10^{33}$ erg s$^{-1}$, and $L_X\sim4\times10^{30}$ erg s$^{-1}$. 
\citet{takata21} did not report what fraction of $L_X$ is produced at the spin period of AR Sco versus in the wind impact site, but the contribution at the spin period seems to be roughly half. This means that if AR Sco was an isolated white dwarf pulsar, its X-ray luminosity would be $L_X\sim2\times10^{30}$ erg s$^{-1}$, which is consistent with $L_x \sim 10^{-3} L_{sd}$. 

For  J0901$-$4046, the corresponding numbers are   $L_{sd}$ of $10^{28}$ erg s$^{-1}$, its X-ray luminosity should be $L_X\sim10^{25}$ erg s$^{-1}$,  if it is a neutron star. On
the other hand, if J0901$-$4046 is a white dwarf pulsar, its spin down luminosity of $L_{sd} \sim 10^{33} $ erg  s$^{-1}$ implies $L_{X}\sim 10^{30}$ erg s$^{-1}$, which would be detectable in X-rays with Chandra.

 \citet{2022NatAs...6..828C} obtained 7 ks of Swift X-ray observations of J0901$-$4046 and derived a $3\sigma$ upper limit to the 0.5-10 keV X-ray flux of $F_X < 1.2\times10^{-13}$ erg cm$^{-2}$ s$^{-1}$, and $L_{X} < 3.2 \times 10^{30}$ erg s$^{-1}$ for a distance of 467 pc.  (They used rather different spectral assumptions than we do, see below.) Hence, the Swift
 observations were not deep enough to detect a potential white dwarf pulsar in this system.

\section{Chandra Observations of PSR J0901$-$4046}

The (presumed \NS) radio pulsar J0901–4046 \citep{2022NatAs...6..828C} has a spin period of $P = 75.89$ s, period derivative
$\dot{P}= 2.25 \times 10^{-13}$, and thus an inferred   dipole
magnetic field  $B = 2.6 \times  10^{14}$ G.  The corresponding spin-down luminosity is $L_{sd} = 2.0 \times 10^{28} $  erg s$^{-1}$. \citet{2022NatAs...6..828C} inferred a distance of   328 to 467 pc from its measured dispersion measure of 52$\pm1$ pc cm$^{-3}$, using the two alternative Galactic free electron density models of \citet{Yao2017ApJ...835...29Y} or \citet{Cordes2002astro.ph..7156C}.

We obtained two Chandra X-ray Observatory \citep{weisskopf02} observations of PSR J0901$-$4046, 24.7 ks on 2025 March 13 and 16.7 ks on 2025 August 25. We placed ACIS-S at
the focus, in Very Faint mode. We extracted counts and spectra from 1.5 arcsec radii at the location of the pulsar, but detect no photons within a 1.5
arcsec error circle of the pulsar in 0.5-8 keV. We use this to estimate an upper limit on the X-ray luminosity.

\citet{gehrels86} gives a 99\% confidence upper limit of 4.6 counts (per time interval) as the average, if 0 counts are detected in that interval. Or we can use a 95\% confidence upper limit of 3.0 counts for 0 counts detected. For a combined exposure time of 41413 seconds, these give
upper limits on count rates of $1.1\times10^{-4}$ or $7.2\times10^{-5}$ cts s$^{-1}$. 
Assuming a $\Gamma=2$ power law, and low ($<10^{21}$ cm$^{-2}$) hydrogen
column density N$_{\rm H}$, these give upper limits in 0.5-10 keV flux of 
$F_x<2.7\times10^{-15}$, or 
$1.8\times10^{-15}$, erg cm$^{-2}$ s$^{-1}$ (for the 99\% and 95\% confidence upper limits, respectively). 
A 200 eV blackbody similarly gives 
$F_x<4.2\times10^{-15}$, or 
$<2.8\times10^{-15}$ erg cm$^{-2}$ s$^{-1}$.  These spectral models span the range of observed millisecond pulsar spectra \citep{Bogdanov06b}, while the $N_H$ estimate of $<10^{21}$ cm$^{-2}$ is justified by its close proximity to the Sun. 

For an upper limit distance of 467 pc, these estimates give $L_x$(0.5-10 keV) $< 7.0\times10^{28}$ (PL, 99\% conf), $<4.7\times10^{28}$ (PL, 95\% conf), $1.1\times10^{29}$ (BB, 99\% conf), or $7.3\times10^{28}$ erg s$^{-1}$ (BB, 95\% conf). If the distance is underestimated by a factor of 3, then these upper limits increase by a factor of 9. 
These limits are all well below the projected $L_X$ for a white
dwarf pulsar, even if the distance is substantially underestimated (unlikely given the proximity of the pulsar, which places it in a well-sampled part of the free electron density models).   Overall, our upper limit is $\sim 50$ times lower than that of \cite{2022NatAs...6..828C}.

\section{Discussion}

Our non-detection of X-rays disfavors the  isolated WD interpretation  of  J0901–4046. This is also consistent with the optical non-detection of a WD in this system. \citet{2022NatAs...6..828C} obtained optical photometry of the field around PSR J0901$-$4046 with the SAAO 1-m telescope, and ruled
out optical counterparts brighter than 20 - 21 mag. The DECam Plane Survey provided $grizY$ band imaging of the same field, which rules out
optical counterparts brighter than 23.5 mag in $g$ at the $5\sigma$ level \citep{saydjari23}. However, given the relatively high extinction of $E(B-V)=0.73$ mag towards PSR J0901$-$4046 and the fact that distances derived from the dispersion measure can be off by a factor of 2-3 \citep{2009ApJ...701.1243D}, it would be possible to hide white dwarfs more massive than about $1.1~M_\odot$ under the worst case scenario (a $3\times$ further distance of $\sim1.5$ kpc).  
Regardless of these issues, the newly obtained  Chandra data favor the neutron star, rather than the white dwarf, interpretation of  PSR J0901-4046.
A binary  WD is also excluded by the phase-connected timing reported by \citet{2022NatAs...6..828C}.

Our results further deepen the mystery of long period transients. The key question is the origin of  high energy  particles that produce radio emission, and  the origin of the electric potential that accelerates these particles. Isolated \NSs\ self-produce  these particles in vacuum breakdown using rotationally-induced EMF, while in binary systems particles are injected from the companion's wind, while acceleration relies, eventually, on the EMF of the wind  \citep[\eg Jovian hectometric radio emission,][]{2014P&SS...99..136H}. In the case of isolated  \NS\  PSR J0901-4046, both these mechanisms are excluded. A remaining possibility is a Solar-flare paradigm, whereby the required accelerating potential is achieved in magnetic reconnection  events in magnetar \mss\ \citep[not rotationally-powered,][]{2002ApJ...580L..65L,tlk,2003MNRAS.346..540L,2017ARA&A..55..261K}.

Magnetars' radio emission clearly shows  transient properties \citep{2006Natur.442..892C}. Thus, magnetically-powered radio emission has a clear prediction: LPTs should show  transient radio emission as the reconnection sites in the \ms\ are activated.

\begin{acknowledgments}
We thank Matt Route for comments on the manuscript.
This research was funded by Chandra Award No. GO4-25005A and GO4-25005B. MK is grateful for support by the
NSF under grant  AST-2508429, and  NASA under grants 80NSSC22K0479, 80NSSC24K0380, and 80NSSC24K0436. COH is supported by NSERC Discovery Grant RGPIN-2023-04264.

\end{acknowledgments}


\bibliography{./0901-23-HST.bib,./BibTex.bib,./citationYuanmingWang.bib}

\bibliographystyle{aasjournal}

\end{document}